# Anomalous Transport in Sketched Nanostructures at the LaAlO$_3$/SrTiO$_3$ Interface

Guanglei Cheng[1], Joshua P. Veazey[1], Patrick Irvin[1], Cheng Cen[1*], Daniela F. Bogorin[1**], Feng Bi[1], Mengchen Huang[1], Shicheng Lu[1], Chung-Wung Bark[2], Sangwoo Ryu[2], Kwang-Hwan Cho[2], Chang-Beom Eom[2] and Jeremy Levy[1#]

[1]*Department of Physics and Astronomy, University of Pittsburgh, Pittsburgh, Pennsylvania 15260, USA;* [2]*Department of Materials Science and Engineering, University of Wisconsin-Madison, Madison, Wisconsin 53706, USA*
\* Present address: Department of Physics, West Virginia University; \*\* Present address: Oak Ridge National Laboratory;  # jlevy@pitt.edu

The oxide heterostructure LaAlO$_3$/SrTiO$_3$ supports a two-dimensional electron liquid with a variety of competing phases including magnetism, superconductivity and weak antilocalization due to Rashba spin-orbit coupling. Further confinement of this 2D electron liquid to the quasi-one-dimensional regime can provide insight into the underlying physics of this system and reveal new behavior. Here we describe magnetotransport experiments on narrow LaAlO$_3$/SrTiO$_3$ structures created by a conductive atomic force microscope lithography technique. Four-terminal local transport measurements on ~10-nm-wide Hall bar structures yield longitudinal resistances that are comparable to the resistance quantum $h/e^2$ and independent of the channel length. Large nonlocal resistances (as large as $10^4\,\Omega$) are observed in some but not all structures with separations between current and voltage that are large compared to the 2D mean-free path. The nonlocal transport is strongly suppressed by the onset of superconductivity below ~200 mK. The origin of these anomalous transport signatures is not understood, but may arise from coherent transport defined by strong spin-orbit coupling and/or magnetic interactions.

## I. Introduction

The discovery of a two-dimensional electron liquid (2DEL) [1] at the interface between two insulating oxides, LaAlO$_3$ (LAO) and SrTiO$_3$ (STO), has attracted considerable attention due to the wide range of properties exhibited. These phenomena emerge from the conducting state of the 2DEL, above a sharp metal-insulator transition (MIT), at a critical LAO thickness of 3 unit cells [2]. The MIT itself is quite remarkable and exhibits a temperature and electric field dependence that is distinct from ordinary semiconductors [3]. The existence of a conducting interface has been explained by a "polar catastrophe" model [1] in which the 2DEL forms to help screen the polarization of the LAO layer.

Though LAO and STO are both non-magnetic, their interface exhibits emergent magnetic behavior [4, 5] in addition to the conducting properties previously reported. Under some conditions, the magnetism can coexist [6-8] with interfacial superconductivity [9, 10]. Strong Rashba spin-orbit coupling, with spin splitting as large as 10 meV, has been inferred from weak antilocalization measurements [11, 12]. The large Rashba interaction is believed to originate from atomic spin-orbit coupling of Ti 3$d$ $t_{2g}$ orbitals combined with inversion symmetry breaking at the LAO/STO interface [13].

The MIT of 3-unit-cell LAO/STO heterostructures can be controlled with extreme nanoscale precision [14] using a voltage-biased conductive atomic force microscope (c-AFM) probe in contact with the top LAO surface. It is possible to change the conducting state to form structures with line widths as narrow as 2 nm and a point resolution approaching 1 nm [15]. This technique has enabled the creation of a variety of nanostructures and devices including nanoscale transistors [15], rectifying junctions [16], optical photodetectors [17], and single-electron transistors [18]. The experiments described here are performed on LAO/STO Hall-cross nanostructures characterized by low-temperature magnetotransport measurements.

**II, Device Preparation**

**A, Sample Growth and Processing**

LAO/STO heterostructures are grown in a similar fashion as in Refs. [18, 19]. The STO substrate is $TiO_2$-terminated by etching in buffered HF for 60 seconds. An atomically smooth surface is achieved by subsequent annealing at 1000°C for 2-12 hours. After substrate preparation, a thin (3.4 unit cells) LAO film is grown on top of STO by pulsed laser deposition at a temperature of either 780 °C or 550 °C. Table 1 summarizes device names and sample growth parameters.

Electrical contacts are made to the LAO/STO interface by $Ar^+$ etching (25 nm) followed by Ti/Au (2 nm/23 nm) sputter deposition. A second Au layer defines wire bonding pads and their leads, connecting them to the contacts from the first Au layer for each "canvas". Electrical contact is made between this second layer and a chip carrier via gold wire bonds.

**B, Device Writing**

Nanoscale devices are written using c-AFM lithography [14]. Prior to device writing, the entire canvas is erased by performing multiple raster scans in contact mode with a tip voltage $V_{tip}$=-10 V. The background resistance between adjacent electrodes (>1 GΩ at room temperature) increases by several orders of magnitude at low temperatures, eliminating the possibility of unwanted coupling between electrodes.

Five-terminal Hall devices with center channel lengths varying between 2 μm and 10 μm are written with positive voltages ($V_{tip}$ =10~15 V) applied to the tip and in a speed of 200 nm/s.

The resulting nanostructures are composed of nanowire segments having width ~10 nm, as determined by analysis of local erasure experiments performed on the same canvas with the same AFM probe [14]. Trapezoidal virtual electrodes contact the gold interface electrodes at one end and taper down to contact the nanostructure at the other end, as shown in Fig. 1(a,b). Upon completion of device-writing, the sample is quickly transferred to the vacuum environment of a dilution refrigerator for cryogenic measurements.

**III, Experiments**

Transport experiments are carried out at two locations, the University of Pittsburgh (Pitt, Device C and H-H7) and the National High Magnetic Field Laboratory (NHMFL, Device C2 and C3), using independent measurement setups. At both locations, true differential voltage amplifiers (Femto model DLPVA) with 1 TΩ input impedance are used so that virtually no current is drawn in the voltage sensing contacts. At Pitt, simultaneous-sampling 24-bit data acquisition cards are used to source voltages and acquire amplified voltage signals. Lock-in amplification is achieved by demodulating the signal with computer-based digital signal processing, allowing simultaneous measurement of up to 4 channels. For electrical current measurements, a transimpedance amplifier is used (Femto DDPCA-300). At NHMFL, conventional lock-in amplifiers are employed to perform transport measurements. Low-frequency (~1 Hz) lock-in detection is used at both locations to minimize capacitive and inductive coupling. Devices are cooled in refrigerators with base temperatures 50 mK (Pitt), 20 mK (NHMFL) or 250 mK (NHMFL).

**IV, Results**

Transport experiments are performed on multi-terminal Hall-cross structures with $W\sim 10$ nm-wide wire segments. All of the results reported here have been reproduced qualitatively in multiple devices from different LAO/STO heterostructures, independent of growth conditions (Table 1) and measurement setups (at Pitt and at NHMFL). Although the devices are fabricated in a nominally similar fashion, two distinct classes of physical properties emerge across all experiments. Of the ten devices reported here, three exhibit "C type" behavior, while the remaining seven exhibit "H type" behavior. The devices themselves have been labeled after measurement according to their classification (e.g., Device C2, or Device H3). For clarity of comparison, an example of each type, Device C and Device H, are shown in Fig. 1 with device schematics and transport characteristics. A detailed summary of device behaviors is shown in Table 1, with additional data on device C2 and C3 given in the supplemental materials (Fig. S1 and S3).

Four-terminal $V^{cd}(I^{ab})$ curves are acquired by sourcing current $I^{ab}$ from lead $a$ to $b$ and simultaneously measuring the voltage $V^{cd}$ between leads $c$ and $d$. The differential resistance $R^{ab,cd} \equiv dV^{cd}/dI^{ab}$ is obtained by numerical differentiation or directly measured with a lock-in amplifier. Figure 1(c) shows linear current-voltage (*I-V*) curves for Device C and H, with four-terminal resistances $R_C^{14,23} = 28.4$ k$\Omega \approx 1.1\ h/e^2$ (at $T$=400 mK) and $R_H^{14,23} = 6.9$ k$\Omega \approx 0.3\ h/e^2$ (at 500 mK), respectively. The longitudinal resistance for both devices is approximately temperature-independent (within a factor of two) up to ~50 K.

Figure 1(d) shows the *nonlocal* transport characteristics of Device C, obtained by sourcing current $I^{12}$ and measuring the resulting voltage across terminals which are separated from the main current channel by 10 μm (e.g., $V^{43}$). The *I-V* relationship is linear with nonlocal resistance $R_C^{12,43} = 2.5\text{ k}\Omega$. Based on low sourcing current as well as the linearity of the nonlocal *I-V* curves, thermoelectric effects can be ruled out; furthermore, inductive couplings are not present since the measurements are performed at quasi-DC frequencies. The nonlocal resistance for Device H (channel length 2 µm) is vanishingly small within experimental measurement limits.

The out-of-plane magnetic field dependence of the nonlocal resistance $R_C^{12,43}$ for device C is shown in Fig. 2(a). The nonlocal magnetoresistance is empirically described by the expression $R_C^{14,23} \propto \sqrt{B_{NL}^2 + B^2}$ where $B_{NL} = 15\text{ T}$. The forward and reverse field sweeps exhibit some hysteresis. In addition, local increases near +/- 3 T are observed after sweeping through zero magnetic field, somewhat similar in nature to the hysteretic effects reported by Brinkman *et al.* [4].The local two terminal magnetoresistance shows overall positive curvature as well as magnetic hysteresis (Fig. 2(b)), with a slow time scale response. Device C3 shows evidence of weak antilocalization (Fig. S3) but without any signature of magnetic hysteresis.

A striking interaction between superconductivity and nonlocal resistance occurs below the superconducting critical temperature $T_c \approx 200\text{ mK}$. Figure 3(a) shows the four-terminal longitudinal $V^{23}(I^{14})$ relationship for Device C at *T*=65 mK, well below $T_c$. As will be described in greater detail elsewhere [20], the low-resistance state for $|I^{14}| < I_c^{14} = 3\text{ nA}$ results from superconductivity within the nanowire. Application of a magnetic field larger than $H_c \approx 1500$ Oe destroys the superconducting state, thereby restoring a linear *I-V* curve. Figure 3(b) shows a nonlocal experiment performed by sourcing current between leads 1 and 2 and measuring the voltage across terminals 3 and 4. The nonlocal resistance collapses with the onset of superconductivity in the main channel. When the driving current or applied magnetic field becomes sufficiently large, superconductivity is suppressed and the nonlocal resistance returns. The nonlocal critical current $I_c^{12} \approx 25$ nA is approximately one order of magnitude larger than $I_c^{14}$. Suppression of the nonlocal response is not directly related to supercurrent flow within the main channel, since there is no current flow within the main channel. Figure 4 shows the nonlocal resistance as a function of current $I^{12}$ and magnetic field $H$. As the magnetic field strength is increased, the nonlocal critical current decreases monotonically. The abrupt increase in nonlocal resistance observed at ~70 nA exhibits a much more rapid decrease with magnetic field strength compared to the main jump at 25 nA. Device H becomes superconducting (Fig. 3(c)) with a similar critical current, but the nonlocal response remains vanishingly small for all values of the magnetic field (Fig. 3(d)).

Figure 5 shows the two-terminal $I$-$V$ curve and its correlation with the nonlocal differential resistance (numerically calculated from Fig. 3(b)) for Device C. The two-terminal $I$-$V$ curve is nonlinear, showing an overall s-shaped profile (higher resistance at lower current) and step-like structures in the range -25 nA $< I^{12} <$ 25 nA. Some sharp spikes, as indicated in arrows in Fig. 5(b), are superconductivity related features that are suppressed by magnetic fields (Fig. 4). Such spikes are strongly correlated between two terminal resistance and nonlocal resistance, further suggesting superconductivity can globally affect the transport properties in the device. Other non-superconductivity related local variations do not affect the nonlocal transport. For example, in the range -150 nA $< I^{12} <$ -50 nA, the two terminal resistance changes from $2h/e^2$ to $3h/e^2$ while the nonlocal resistance remains unchanged.

The contact resistance can be estimated from the difference of the two terminal and four terminal resistance measurements. In the normal state, the four-terminal resistance $R_c^{14,23}=1.1h/e^2$, while the two-terminal resistance $R_c^{12,12}$ ~ 2-3 $h/e^2$. These values fall significantly below the input impedance of the voltage amplifiers or any other characteristic impedance in the system.

The length dependence of longitudinal resistances is examined for all type C and H devices (i.e., those with no observable nonlocal signal) at 500 mK. As shown in Fig. 6, while the length varies from 2 to 10 μm and for all the 10 nm wide H type devices, the longitudinal resistances of these five devices fall within a relatively narrow range 6.5-8.6 k$\Omega$. Type-C devices show a wider variation, but no clear scaling with main channel length.

**V, Discussion**

The local and nonlocal transport phenomena reported here place strong constraints on any unified theoretical description. The 3 u.c. LAO/STO system is naturally insulating—sheet resistances of erased canvases are of order GΩ at room temperature and immeasurably large at cryogenic temperatures. Thermoelectric effects cannot account for the observation of the large nonlocal signal, since the *I-V* curve would rather have a quadratic or nonlinear relationship than the linear nonlocal *I-V* curve (Fig. 1(d)) observed here. The absence of charge-current flow along the main channel implies the existence of another state variable to relate voltages and currents, e.g., spin degree of freedom. A key observation is that the nonlocal transport of type-C devices spans the device dimensions (*L*~10 μm), two orders of magnitude larger than the mean-free path ($l_{MF}$~10-100 nm) of 2D LAO/STO [21].

There are a number of reported nonlocal transport effects in semiconductors and superconductors that one can look toward in search of a theoretical explanation of these observations. Superconductivity-related nonlocal effects, e.g. Andreev reflection and charge imbalance [22], cannot account for the presence of nonlocal transport well above the superconducting critical temperature, e.g., at 4 K. Large nonlocal transport in observed in the quantum spin Hall insulator phase of HgTe/CdTe quantum wells [23-25]. However, the electron bands in LAO/STO are all electron-like, so it is unclear how a bulk energy gap could arise. The nonlocal resistances observed in spin Hall effect experiments [26] are generally too small to observe without a spin-polarized source of electrons [26, 27] or a highly sensitive detector [28]. Nonlocal transport associated with a large spin/valley Hall effect has been observed near the Dirac point in graphene [29], however, a strong magnetic field is needed to observe this effect.

The experimental results suggest some form of highly coherent charge and/or spin transport. As shown in Fig.6, the longitudinal resistance of type-H devices is very weakly correlated with channel length, varying by only 20% as the channel length is changed by 500%. For type-C devices, C and C2 have similar longitudinal resistances despite the channel length differs from 6.5 to 10 μm. C3 have much smaller nonlocal and longitudinal resistances, possibly suggesting the presence of a conducting "bulk" within the nanowire that screens the nonlocal resistance and contributes to the longitudinal resistance. Additionally, the resistance values are of order $h/e^2$ for all of the devices. In the mean time, our previous results reveal anomalous high mobilities that persist to room temperatures in these nanowires [30].

While the origin of the nonlocal transport is an open question, we briefly discuss a few possibilities. The presence of nonlocal transport implies that nonlocal voltages cannot be screened. If the nanowire should be regarded as a 2D system, then the bulk conduction states would need to be gapped. Magnetoconductance oscillations in 2D LAO/STO devices have been observed as a function of in-plane magnetic fields, which have been attributed to the opening and closing of a spin-orbit gap [31]. Lateral quantum confinement or disorder [32, 33] can also open a gap at the Fermi energy.

Another possible explanation for the nonlocal transport, proposed by Fidkowski *et al.* [34], involves the possible existence of a novel helical band structure in the 1D LAO/STO nanowires. Ferromagnetic exchange between localized and delocalized carriers produces spontaneous ferromagnetism, and strong spin-orbit interaction leads to a helical band structure in which spin and momentum are locked. Such a helical wire could support nonlocal transport in the presence of spin imbalance due to coupling of spin-motion locked transmission modes to the magnetization in different leads.

The physical distinction between type-C and type-H devices is not understood, but several factors are likely to be involved. First, it is known that LAO/STO samples exhibit local micron-scale inhomogeneous magnetic patches, as has been observed from scanning SQUID measurements [7]; Such magnetic inhomogeneity is uncontrollable and the physical origin is still under debate. However, qualitative differences between the devices could be linked to this inhomogeneous magnetism. In fact, one of the devices (Device C) explicitly shows ferromagnetic behavior in one of the leads (Fig. 2(b)). Second, type-H devices may lack the same ferromagnetic coupling. Alternatively, they may also have a Fermi energy that locally lies outside of the spin-orbit gap window due to the variations of carrier density or confinement depth along the wire. Such variations are expected due to the nature of the c-AFM writing technique. Experiments on wider structures (H6 and H7) show type-H behavior, i.e., absence of nonlocal signals.

## VI. Conclusions

Transport experiments have been performed on nanoscale devices created at the LaAlO$_3$/SrTiO$_3$ interface. Local transport experiments show longitudinal resistance values comparable to the resistance quantum and the existence of large (~kΩ) nonlocal transport signatures over ~10 μm scales that greatly exceed the 2D mean free path. The nonlocal transport is strongly suppressed by the onset of superconductivity below ~200 mK. The unusual transport behavior suggests a spin-based mechanism related to emergent magnetism and strong spin-orbit coupling.

**Acknowledgements**

We thank Lukasz Fidkowski, Sergey Frolov, Xiaopeng Li, Roman Lutchyn, Chetan Nayak, and Di Xiao for stimulating discussions, and helpful comments on the manuscript. This work is supported by AFOSR FA9550-10-1-0524 (J.L., C.B.E.), ARO W911NF-08-1-0317 (J.L.), NSF DMR-1104191 (J.L.), and DMR-0906443 (C.B.E). A portion of this work was performed at the National High Magnetic Field Laboratory, which is supported under NSF Cooperative Agreement No. DMR-0654118, No. DMR-1157490, the state of Florida and the U.S. Department of Energy.

Table 1. Device growth parameters and transport characteristics.

| Device Name | Growth Temperature (°C) | Oxygen Pressure (mBar) | Annealing Conditions | Channel length ($\mu$m) | Channel width (nm) | Normal state Nonlocal resistance ($\Omega$) |
|---|---|---|---|---|---|---|
| **C**  | 780 | $7.5\times10^{-5}$ | 600°C | 10  | 10  | 2500 |
| **C2** | 550 | $1\times10^{-3}$   | No    | 6.5 | 10  | 1500 |
| **C3** | 550 | $1\times10^{-3}$   | No    | 12  | 10  | 20 |
| **H**  | 550 | $1\times10^{-3}$   | No    | 2   | 10  | 0 |
| **H2** | 780 | $7.5\times10^{-5}$ | 600°C | 10  | 10  | 0 |
| **H3** | 550 | $1\times10^{-3}$   | No    | 6.5 | 10  | 0 |
| **H4** | 550 | $1\times10^{-3}$   | No    | 6.5 | 10  | 0 |
| **H5** | 550 | $1\times10^{-3}$   | No    | 10  | 10  | 0 |
| **H6** | 550 | $1\times10^{-3}$   | No    | 2   | 250 | 0 |
| **H7** | 550 | $1\times10^{-3}$   | No    | 2   | 50  | 0 |

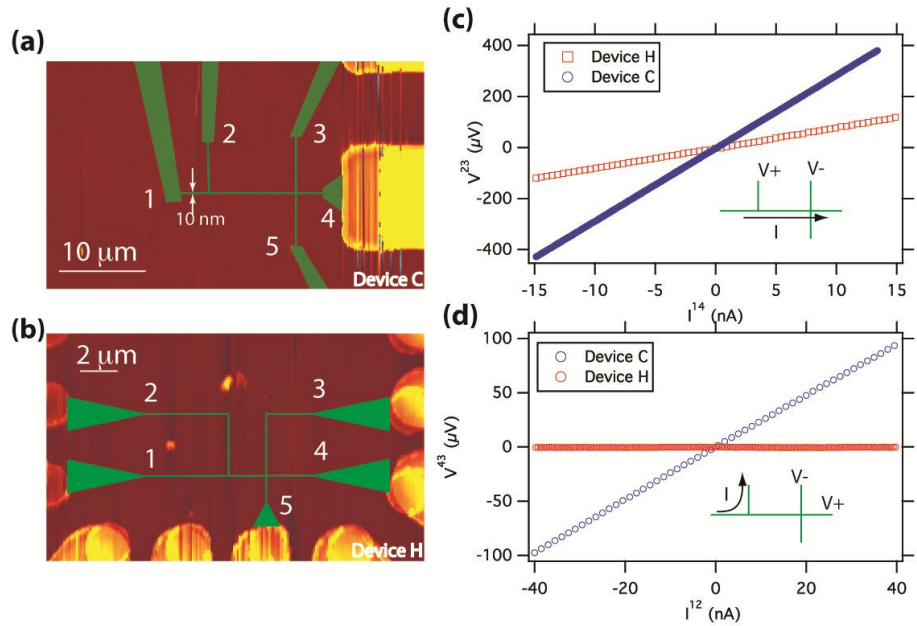

FIG. 1. Device schematic and electrical characterization of LAO/STO nanostructures. (a) Five-terminal structure (Device C, overlaid on an AFM canvas image) composed of $W$=10 nm wide conducting nanowires (green), including an $L$=10 µm long main channel surrounded by insulating (red) background. (b) Five-terminal nanowire (Device H) formed in a similar fashion ($W$=10 nm, $L$=2 µm). (c) Longitudinal *I-V* curves for Devices C and H at $T$=400 mK. Inset shows direction of current flow and voltages measured. (d) Nonlocal *I-V* curves for Devices C and H at $T$=400 mK. Inset shows direction of current flow and voltages measured.

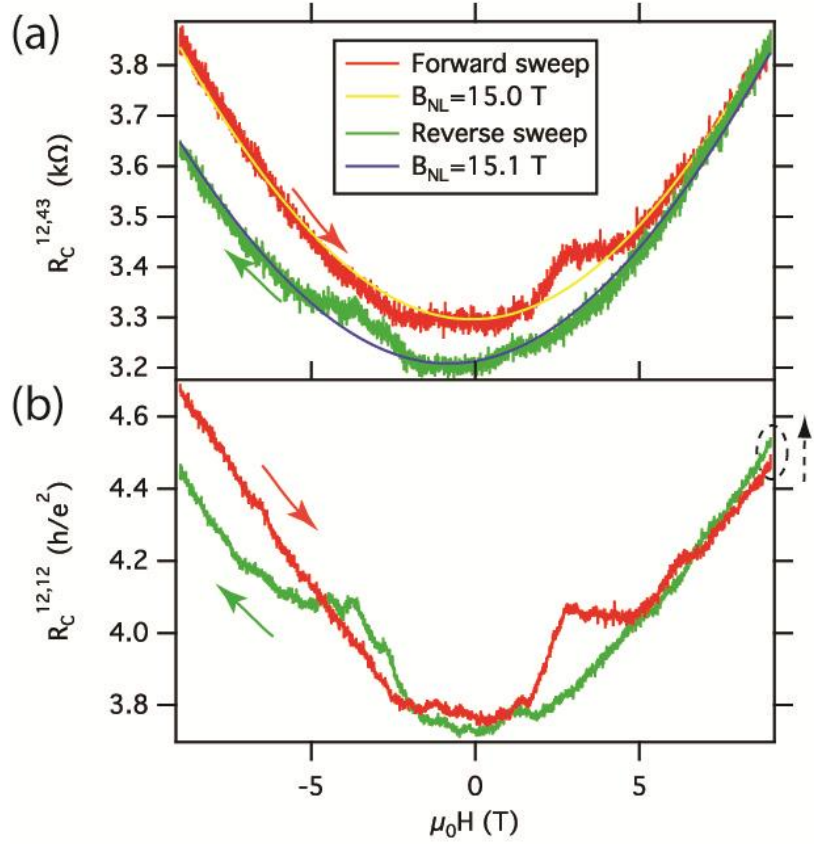

FIG. 2. Out-of-plane field dependence of (a) nonlocal resistance $R_C^{12,43}$ and (b) local two terminal resistance $R_C^{12,12}$ at $T$=4 K. Red and green curves represent forward and reverse sweeps (at 3 mT/s). Yellow and blue solid lines in (a) are empirical fits with form $R_C^{14,23} \propto \sqrt{B_{NL}^2 + B^2}$, giving two similar built-in fields $B_{NL}$ of 15.0 T and 15.1 T.

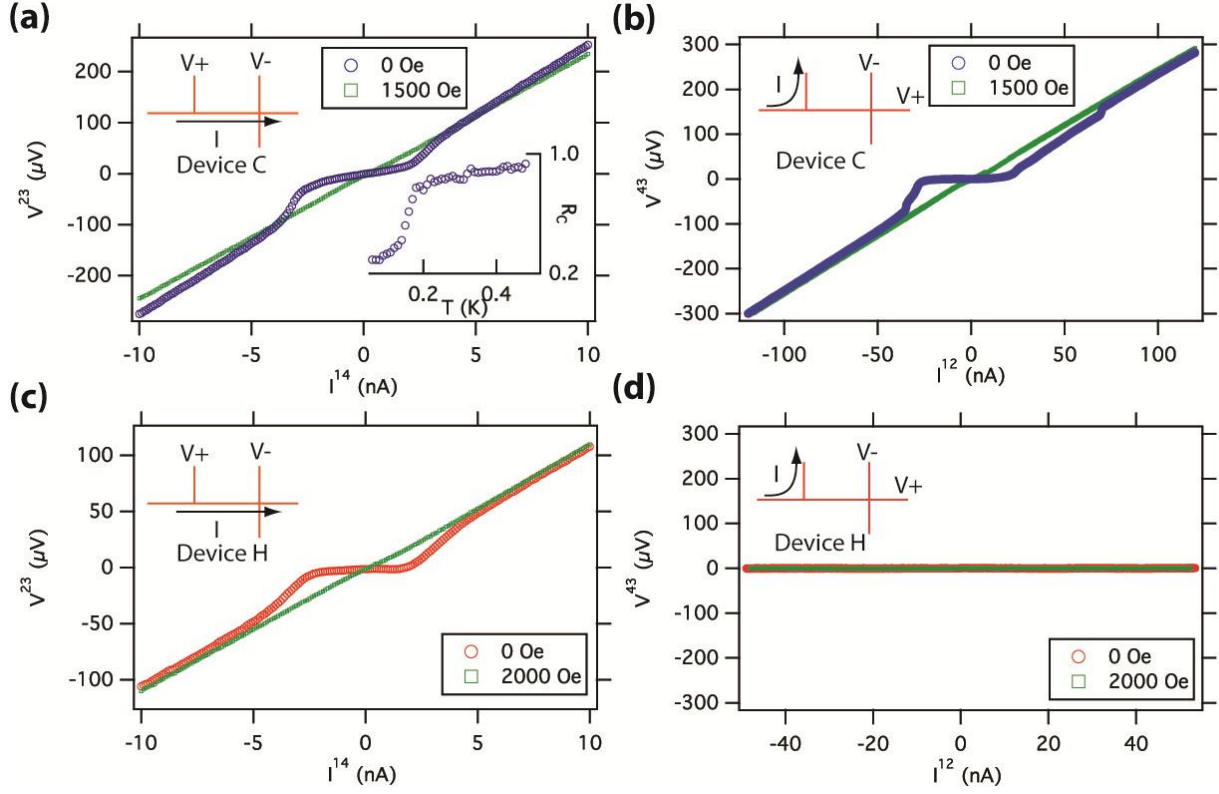

FIG. 3. Interaction between Cooper pairs and edge states. (a) Longitudinal *I-V* curves for Device C at *T*=65 mK, below the superconducting critical temperature. The lower-resistance region of the H=0 curve is associated with the formation of Cooper pairs. Normal-state conduction is restored with application of an external magnetic field (green curve). The inset shows the temperature dependence of longitudinal resistance, indicating the critical temperature is ~200 mK. Y axis unit is $h/e^2$. (b) Nonlocal resistance $R_{12,43}$ of Device C under conditions where Cooper pairs form (blue curve). The Cooper pairs block spin transport without current flow that is required for manifestation of nonlocal resistance. Application of an external magnetic field restores the linear nonlocal response (green curve). (c) Longitudinal *I-V* curves for Device H (*T*=50 mK) in the superconducting regime (*H*=0) and normal state (*H*=2000 Oe). (d) Nonlocal resistance $R_C^{12,43}$ of Device H under conditions where Cooper pairs form (*T*=50 mK). The nonlocal signal is vanishing both above and below $T_c$.

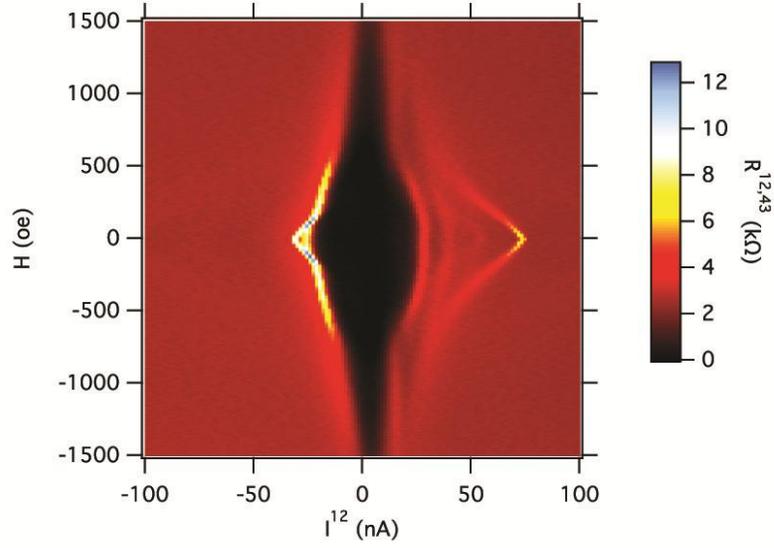

FIG. 4. Nonlocal differential resistance dependence of driving current and external out-of-plane magnetic field for Device C. Nonlocal resistance is numerically extracted from a series of *I-V* curves similar to Fig. 3(b).

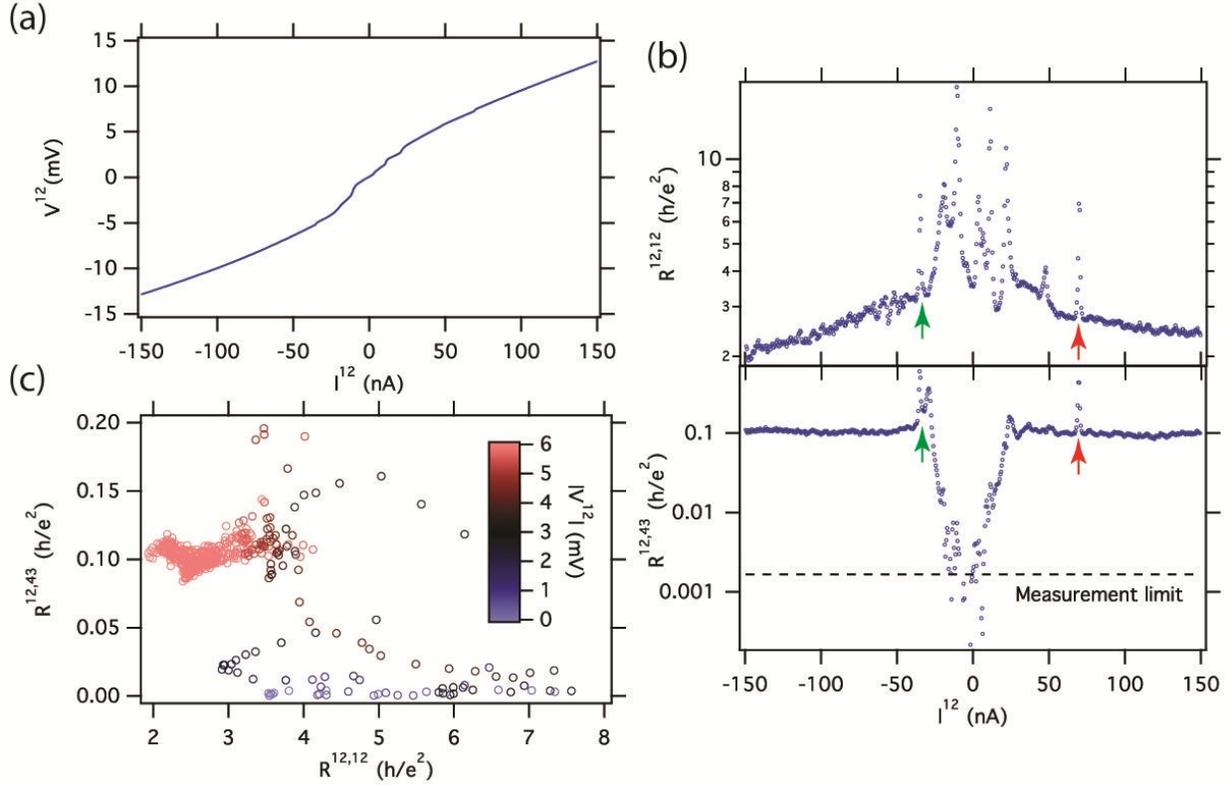

FIG. 5. Correlations between local and nonlocal transport at $T$=70 mK and $H$=0 Oe. (a) Local $I$-$V$ curve showing non-linear response. (b) Local differential resistance (top panel) shows several sharp features and nonlocal resistance (bottom panel) drops by two orders of magnitude at sufficiently small $I^{12}$. There are many correlations between the local transport between leads 1 and 2 and the nonlocal voltages that appear at leads 3 and 4. Such correlations are indicated by gree and red arrows. (c) Parametric plot of local and nonlocal resistance for various current values $I^{12}$. The local resistance is larger when the nonlocal resistance is small, and vice-versa.

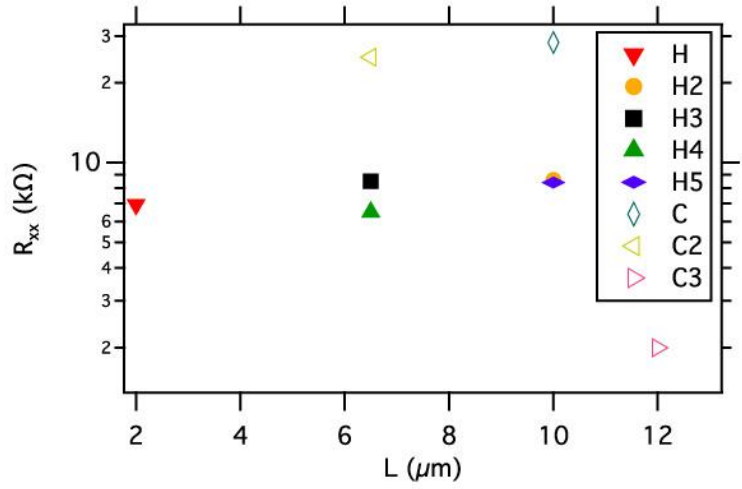

FIG. 6 Four-terminal longitudinal resistance of type-C and type-H devices with various channel lengths, and fixed width of 10 nm at T=500 mK.

# Supplemental Materials for

# "Nonlocal Transport in Sketched Oxide Nanostructures"

Guanglei Cheng[1], Joshua P. Veazey[1], Patrick Irvin[1], Cheng Cen[1,*], Daniela F. Bogorin[1,**],

Feng Bi[1], Mengchen Huang[1], Shicheng Lu[1], Chung-Wung Bark[2], Sangwoo Ryu[2],

Kwang-Hwan Cho[2], Chang-Beom Eom[2] and Jeremy Levy[1,#]

[1]*Department of Physics and Astronomy, University of Pittsburgh, Pittsburgh,*

*Pennsylvania 15260, USA;* [2]*Department of Materials Science and Engineering,*

*University of Wisconsin-Madison, Madison, Wisconsin 53706, USA*

* Present address: Department of Physics, West Virginia University

** Present address: Oak Ridge National Laboratory

# jlevy@pitt.edu

**Devices C2 and H2**

Figure S1 shows data from an additional device (Device C2) that exhibits a similar nonlocal response as Device C. The longitudinal normal resistance for a 6.5 μm section is 25.0 kΩ at $T$=25 mK and 10 kOe out-of-plane magnetic field. This longitudinal resistance ($\sim h/e^2$) is close to that of Device C, which has a 10 μm long channel. The carrier density, $n_{c2} = 1.8 \times 10^{13}$ cm$^{-2}$, is extracted from the Hall resistance vs. magnetic field dependence (Fig. S1(b)). Such a value is lower than Device C, but larger than Device H. The anomalous Hall response gives a zero-field Hall resistance of 30 Ω. Like Device C, a large nonlocal signal (1.6 kΩ) is observed and is suppressed by superconductivity below the critical field in Device C2 (Fig. S1(c)). Above the critical field, superconductivity is suppressed and the nonlocal signal is restored.

Another device (H2) with the same geometry as Device C is shown in Fig. S2. The transport behavior (Fig. S2(b) and S2(c)) is characteristic of the Device H with a vanishing non-local resistance. The normal-state longitudinal resistance is very close to Device H, despite the fact that the length is 5 times longer.

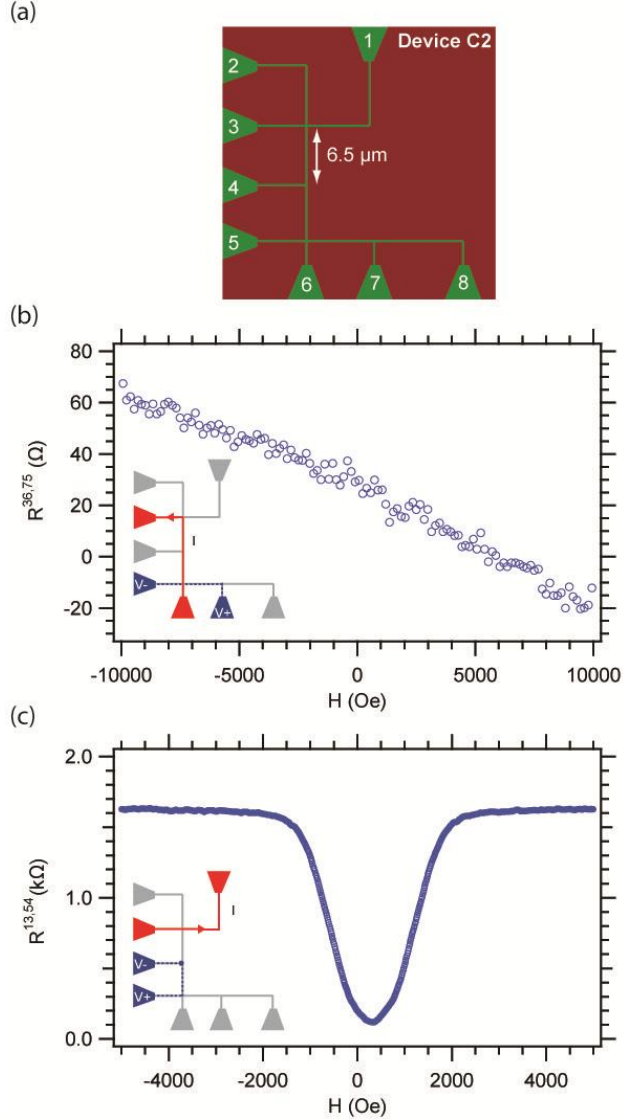

FIG. S1. Device C2. (a) Schematic of Device C2, which has eight leads. The lengths of consecutive segments are 6.5 μm between the pairs 3 & 4, 4 & 5, and 6 & 7. Field-dependent measurements at $T$=18 mK acquired with a lock-in amplifier, with $V_{exc} = 1 \text{ mV}; f = 1 \text{ Hz}$. (b) Hall resistance vs. magnetic field. (c) Nonlocal resistance vs. magnetic field, showing that the nonlocal signal collapses (the dip in the curve) at fields where Cooper pairs are allowed to form. Above a critical field $H_{c2}$, the nonlocal signal is restored. The dip position is slightly shifted off zero due to a hysteretic effect in the magnet.

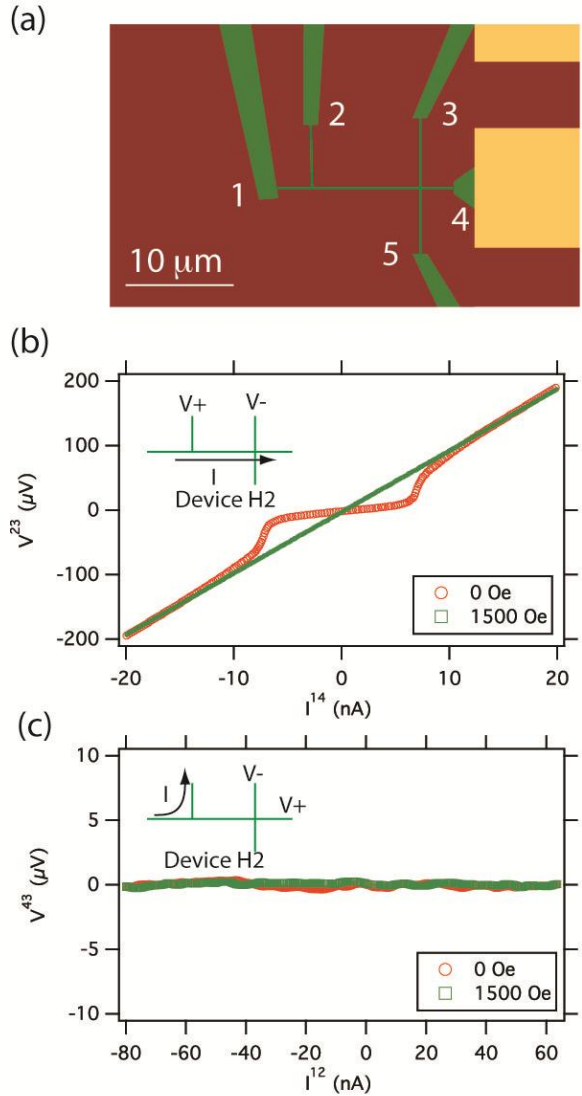

FIG. S2. Device H2 at 50 mK. (a) Schematic of Device H2. (b) Longitudinal *I-V* curves of device H2 at 0 and 1500 Oe external magnetic fields, indicating Cooper pair formation. The normal channel resistance and superconducting channel resistance are 9.6 kΩ and 1.7 kΩ, respectively. (c) Absence of nonlocal signals measured at 0 and 1500 Oe external magnetic fields.

## Device C3

Device C3 is a 6 terminal Hall-bar written in a similar fashion as device C and C2, with a longest main channel of 12 μm. Device was cooled down in a He-3 refrigerator with a base temperature of 250 mK at NHMFL (still above $T_c$). Transport measurement was done through conventional lock-in amplification at an excitation frequency of 7.1 Hz. Figure S3 shows the nonlocal transport data as a function of field and temperature. The nonlocal resistance is suppressed at low magnetic fields and temperatures due to the onset of superconductivity (Fig. S3(b)). Weak antilocalization manifested as cusps around zero field is observed as a function of field orientation, suggesting strong spin-orbit coupling existing in this device (Fig. S3(c)).

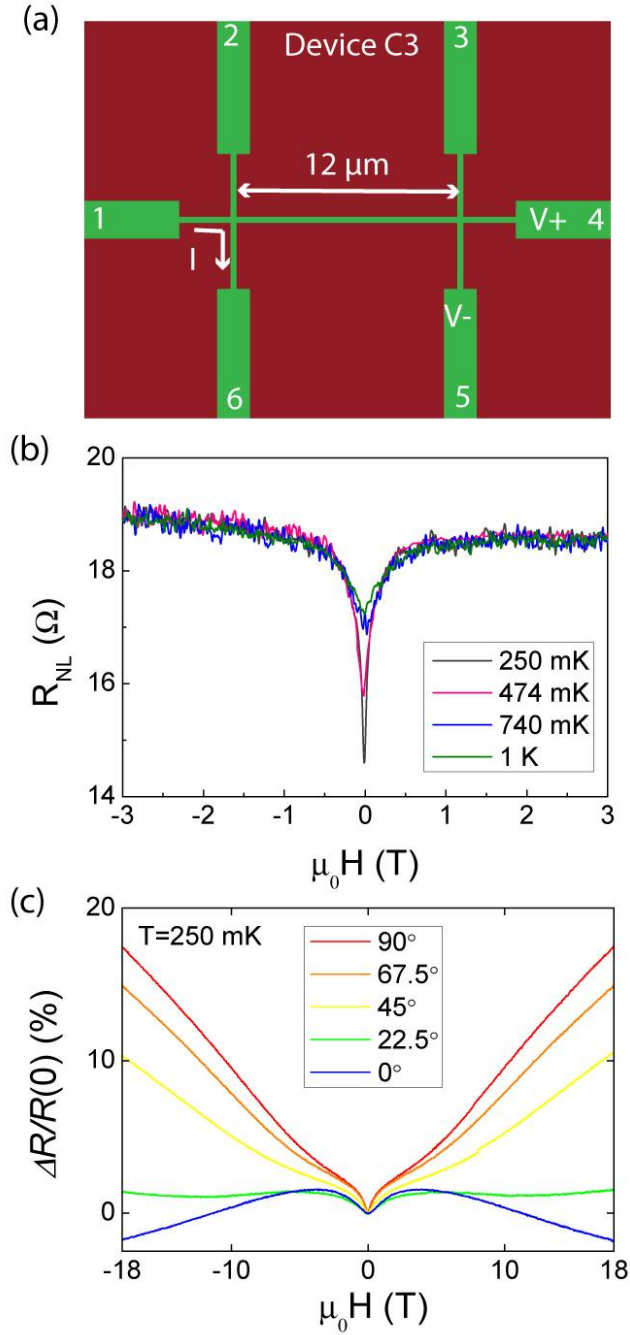

Fig. S3. Device C3. (a) Schematic of device C3. (b) Temperature dependence and field dependence of nonlocal resistance $R_{16,45}$. (c) Local two terminal magnetoresistance $(R_{16,16}(H)-R_{16,16}(0))/R_{16,16}(0)$ as a function of field orientation from in-plane (0 degree) to out-of-plane (90 degrees), showing signatures of weak antilocalization.